\documentclass[notoc,12pt]{JHEP3}

\usepackage{amsmath,amssymb,euscript,array,cite}

\input{epsf}
\newcommand{\startappendix}{
\setcounter{section}{0}
\renewcommand{\thesection}{\Alph{section}}}
\newcommand{\Appendix}[1]{
\refstepcounter{section}
\begin{flushleft}
{\large\bf Appendix \thesection: #1}
\end{flushleft}}

\usepackage{epsfig}

\def\N{{\cal N}}

\def\ttau{{\tilde\tau}}

\def\Tr{{\rm Tr}\,}

\def\det{{\rm det}}
\def\SU{\text{SU}}

\def\U{\text{U}}

\def\Dbarslash{\,\,{\raise.15ex\hbox{/}\mkern-12mu {\bar\D}}}
\def\Dslash{\,\,{\raise.15ex\hbox{/}\mkern-12mu \D}}
\def\delslash{\,\,{\raise.15ex\hbox{/}\mkern-9mu \partial}}
\def\delbarslash{\,\,{\raise.15ex\hbox{/}\mkern-9mu {\bar\partial}}}

\def\PD#1#2{\frac{\partial #1}{\partial #2}}

\newcommand{\EQ}[1]{\begin{equation} #1 \end{equation}}
\newcommand{\AL}[1]{\begin{subequations}\begin{align} #1
\end{align}\end{subequations}}

\title{Five-Dimensional Gauge Theories and Quantum Mechanical Matrix Models}
\author{Timothy J.~Hollowood\\
Department of Physics, University of Wales Swansea,
Swansea, SA2 8PP, UK\\
E-mail: {\tt t.hollowood@swan.ac.uk}}
\preprint{SWAT-}
\abstract{We show how the Dijkgraaf-Vafa matrix model proposal can be
  extended to describe five-dimensional gauge theories compactified on
  a circle to four dimensions. This involves solving a certain quantum
  mechanical matrix model. We do this for the lift of the $\N=1^*$
  theory to five dimensions. We show that the resulting expression for
  the superpotential in the confining vacuum is identical
  with the
  elliptic superpotential approach based on Nekrasov's
  five-dimensional generalization of Seiberg-Witten theory involving
  the relativistic elliptic Calogero-Moser, or Ruijsenaars-Schneider,
  integrable system. 
}

\begin{document}

\section{Introduction}

The Dijkgraaf-Vafa approach for calculating the
superpotential of a four-dimensional supersymmetric gauge theory via a
matrix model \cite{DV3,DV2,DV1} is by now well established. The matrix model
seems capture all the holomorphic information of an $\N=1$ theory in a
lower dimensional---in this case a zero dimensional---system. Recently
Dijkgraaf and Vafa \cite{DV4} have proposed a much more general form of
their approach which would apply, for instance, to higher dimensional
theories compactified to four dimensions. This far-reaching proposal
is very striking and without doubt deserves detailed study and
independent testing.

From the original proposal involving matrix integrals, 
it has turned out that relevant
perturbations of the $\N=4$ theory, the so-called $\N=1^*$ theory, 
\cite{DV1,mm1,mm2}, and its Leigh-Strassler
deformations \cite{cft,lsint}, 
can be solved since the associated matrix models were studied in other
contexts \cite{Kazakov:1998ji,Kostov}. 
For the $\N=1^*$ theory the matrix model results agree with 
the elliptic superpotential approach which is an alternative approach 
for calculating the
holomorphic condensates \cite{nick,us}. In order to investigate the recent
higher-dimensional proposal, it is natural to consider
relevant deformations of the $\N=4^*$ theory lifted to five dimensions
where the additional dimension is compact. The recent work of Dijkgraaf
and Vafa would have us consider a quantum mechanical matrix model rather
than a matrix integral. We shall solve this model in Section
2 and hence compute the superpotential in the confining vacuum of the
theory. The results in the other massive vacua follow by a
straightforward generalization of \cite{mm2}.

Fortunately we have an
independent and highly non-trivial check on the result based on the
elliptic superpotential approach adopted in \cite{nick,us}. The
generalization to the five dimensional theory has not been described previously
but it not difficult to establish the form of the exact elliptic superpotential
in this case, generalizing the one for the $\N=1^*$ theory 
in \cite{nick}. The point is that the
Seiberg-Witten theory of the compactified five-dimensional theory was
considered by Nekrasov \cite{Nek}. What happens is that the
Calogero-Moser integrable system which underlies
the four dimensional theory is replaced by its relativistic version:
the so-called Ruijsenaars-Schneider system \cite{RS}. The exact elliptic
superpotential for the simplest mass deformation of the $\N=2$ theory 
is identified with 
the basic Hamiltonian of this integrable system. Hence we are
able to independently calculate, for instance, the gluino condensate in
the confining vacuum of the compactified five dimensional theory and
the result is entirely consistent with the expression calculated via
matrix quantum mechanics.

As a side remark, we shall find that the quantum-mechanical matrix
model can be reduced to a matrix integral and this integral is very
similar to the one that described the relevant deformations of the 
Leigh-Strassler theory in \cite{cft}. It is rather natural, therefore, that the
elliptic superpotential side of the story should involve the same
integrable system \cite{lsint}; 
namely, that of Ruijsenaars and Schneider. Surely
this is a strong hint of some deeper connection between the integrable
system and the matrix model.

\section{The quantum-mechanical matrix model}

We begin by formulating the superpotential of the compactified
five-dimensional theory as a quantum mechanical matrix model following
Dijkgraaf and Vafa \cite{DV4}. From the point-of-view of the
four-dimensional theory there are 3 adjoint chiral fields $\Phi_i$,
$i=1,2,3$. In the five-dimensional theory, the imaginary component of, say,
$\Phi_3$ is re-interpreted as the component of the gauge field along
the extra compact dimension, while the real part of $\Phi_3$ is
the real scalar of the five-dimensional theory. We will take the
compact dimension to have length $\beta$. The other fields $\Phi_1$
and $\Phi_2$ form an adjoint-valued hypermultiplet. 
The superpotential of the effective 
four-dimensional theory is determined by the quantum mechanical system
involving the fields $\Phi_i(t)$.
The partition function of the quantum mechanical system involves the
functional integral:
\EQ{
Z=\int\prod_{i=1}^3[d\Phi_i]\,\exp\big(-(\beta g_s)^{-1}\int
dt\,W[\Phi_i]\big)\ .
\label{part}
}

The action of the matrix model is a generalization of the one that describes
the $\N=1^*$ deformation of the four dimensional theory:
\EQ{
W[\Phi_i]={\rm
  Tr}\big(i\Phi_1D\Phi_2+m\Phi_1\Phi_2+\mu\cosh(\beta\Phi_3)\big)
\label{act}
}
where the covariant derivative is
$D\Phi_2=\partial_t\Phi_2+[\Phi_3,\Phi_2]$.  
In fact we have been a bit implicit 
in writing down \eqref{part} because we have not 
specified the measure. Part of the Dijkgraaf-Vafa prescription involves
interpreting the matrix integral in a holomorphic way. So the complex
fields $\Phi_1(t)$ and $\Phi_2(t)$ are subject to a particular reality
condition, namely $\Phi_1^\dagger=\Phi_2$, or equivalently
$\Phi_1+\Phi_2$ and $i(\Phi_1-\Phi_2)$ are Hermitian, 
from the point-of-view of the functional
integral. In particular the measure for the latter combination of fields is the
appropriate measure for Hermitian fields. 
The field $\Phi_3(t)$ is treated in a somewhat different
manner as described in \cite{DV4}. First of all,
local gauge transformations can be used to gauge away the
non-constant part of the component of the gauge field along the
compact direction. This leaves large gauge transformations
which shift the eigenvalues of the gauge field by integer multiples of $2\pi
/\beta$. This means that the natural variable is not the gauge field, but
rather its holonomy around the circle $\exp i\beta A_t$. In the
holomorphic point-of-view the quantity $iA_t$ is then
naturally complexified to $\Phi_3$ by including the real scalar and
then the prescription of Dijkgraaf and Vafa \cite{DV4} is 
to interpret the integral as being over the $t$-independent quantity
\EQ{
U=\exp\beta\Phi_3
}
thought of as a unitary matrix. 
Notice that the final term in \eqref{act} 
is the natural generalization of the ${\rm Tr}\,\Phi_3^2$ in the
four-dimensional theory and incorporates the necessary periodicity
$\Phi_3\to\Phi_3+2\pi i/\beta$  
and will allow us to make contact with the generalization of the
condensate $u_2$ described in \cite{Nek}.

Now following the logic of \cite{DV1,mm1}, we integrate out the
fields $\Phi_1(t)$ and
$\Phi_2(t)$ since they appear Gaussian in \eqref{part}. The
complication is that we must now integrate out all the fourier
modes of these fields. We end up with a pure unitary
(zero-dimensional) matrix integral
\EQ{
Z=\int dU\,\frac{\exp\big(-g_s^{-1}
 \mu\,\Tr\cosh(\beta\Phi_3)\big)}{
\det\,\sinh\tfrac12\big(im+\Phi_3\otimes{\bf1}-{\bf1}\otimes\Phi_3\big)}
\ ,
}
where we used the identity
\EQ{
\det(\partial_t+\varphi)
=\det\,\sinh\tfrac12(\beta\varphi)/(\tfrac12\beta)\ .
}

Now we are ready to perform the large-$N$ limit saddle-point
evaluation of the remaining unitary integral around the 
critical point appropriate to the confining vacuum. As usual in a
unitarity matrix integral we can
diagonalize $U$ and work in terms of its eigenvalues
\EQ{
U\thicksim\big(e^{\phi_1},\ldots,e^{\phi_N}\big)
}
at the expense of introducing the unitary matrix version of the
Vandermonde determinant:
\EQ{
Z=\int\prod_id\phi_i\, \prod_{i\neq j}\frac{\sinh\tfrac12(\phi_i-\phi_j)}
{\sinh\tfrac12(\phi_i-\phi_j+i\beta m)}\,
\exp\big(-g_s^{-1}\mu\sum_i\cosh\phi_i\big)\ .
}
Apart from the potential, 
this is precisely the matrix integral that appears in the solution of
the six vertex model on a random lattice \cite{Kostov} and has already been
employed to describe relevant perturbations of Leigh-Strassler
deformations of $\N=4$ theories. However, it is almost identical to
the matrix model that was considered in \cite{Hoppe:1999xg} which
arose from taking a dimensional reduction of $\N=1$ supersymmetric
Yang-Mills in four dimensions to one dimension. The only difference is
that the periodicity in the eigenvalues in that reference is 
along the real axis, however, this is only a superfical difference because our
integral is interpreted in a holomorphic way. In addition, our model
is also identical to the different relevant deformation of the
Leigh-Strassler theory considered in \cite{Mansson:2003dm}.
In order to solve the model, we will largely follow the approach of 
\cite{cft} (or \cite{Mansson:2003dm}) which is tailored towards the
Dijkgraaf-Vafa application. 

In the large-$N$ limit, the eigenvalues $\phi_i$ form a continuum and
condense onto cuts in the complex $z$-plane (actually the cylinder due to the
identification $z\sim z+2\pi i$). One can think of these cuts as arising from
a quantum smearing-out of the classical eigenvalues. For the
confining vacuum all the classical eigenvalues are degenerate $\phi_i=0$
and so we expect a solution in the matrix model involving 
a single cut which we take to extend from $-a$ to $a$. Notice that at
this point we diverge from the Leigh-Strassler case where the cut is
not symmetrical about $z=0$ \cite{cft}. 
The extent of the cut and the matrix
model density of eigenvalues $\rho(z)$ will be determined
self-consistently from the saddle-point 
equation in terms of the 't~Hooft coupling of the matrix model $S=g_sN$. The
saddle-point equation is most conveniently formulated after defining the 
resolvent function
\EQ{
\omega(z)=\tfrac12\int_{-a}^a d\phi\,
{\rho(z)\over\tanh{{z-\phi}\over 2}}\ ,\qquad
\int_{-a}^a\rho(\phi)\,d\phi=1\ .\label{defres}
} 
This function is analytic in $z$ and its only singularity is along a branch
cut extending between $[-a,a]$. The matrix model spectral density
$\rho(\phi)$ is equal to the discontinuity across the cut 
\EQ{
\omega(\phi+i\epsilon)-\omega(\phi-i\epsilon)=-2\pi
i\rho(\phi)\ ,\qquad \phi\in[-a,a]\ .
}
In this, and following equations, $\epsilon$ is an infinitesimal. 
The saddle-point equation expresses the condition of  
zero force on a test
eigenvalue in the presence of the large-$N$ distribution of
eigenvalues along the cut:
\EQ{
\frac{\mu\sinh\phi}{S}=\omega(\phi+i\epsilon)+\omega(\phi-i\epsilon)
-\omega(\phi+i\beta m)-\omega(\phi-i\beta m)\ ,\qquad\phi\in[-a,a]\ .
\label{spe}
}
This equation can be re-written in terms of the useful function
\EQ{
G(z)=\frac{\mu}{2\sin(\tfrac{\beta m}2)}\cosh z+i S\big(
\omega(z+\tfrac{i\beta m}2)-\omega(z-\tfrac{i\beta m}2)\big)\
.\label{defg}
}
{}From this definition, one may easily deduce that $G(z)$ is an analytic
function on the cylinder $-\pi\leq z\leq \pi$, where ${\rm
  Im}\,z=\pi$ and ${\rm Im}\,z=-\pi$ are identified, with two cuts 
$[-a+\tfrac{i\beta m}2,a+\tfrac{i\beta m}2]$ and 
$[-a-\tfrac{i\beta m}2,a-\tfrac{i\beta m}2]$. This is illustrated in
Figure 1. Note that everything is periodic in $\beta m\to\beta m+2\pi$
so we can always choose $-\pi<\beta m<\pi$.  
\begin{figure}
\begin{center}\mbox{\epsfysize=5.5cm\epsfbox{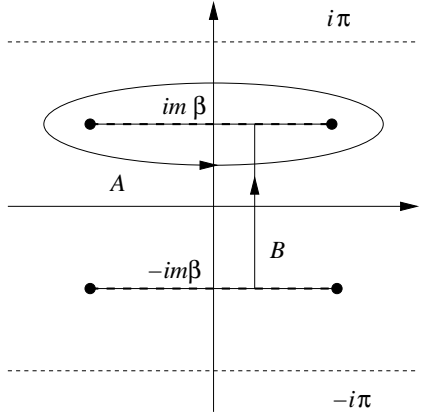}}\end{center}
\caption{\small The region over which the function $G(z)$ is defined
with its two cuts. The lines ${\rm Im}\,z=\pm\pi$ are identified.}
\end{figure}

In terms of $G(z)$, the matrix model saddle-point equation \eqref{spe} is 
\EQ{
G(\phi+ \tfrac{i\beta m}2\pm i\epsilon)=G(\phi-\tfrac{i\beta m}2\mp
i\epsilon)\ ,\qquad\phi\in[-a,a]\ .
\label{glue}
}
This equation can be viewed as a condition which glues the
top (bottom) of the upper cut to the bottom (top) of the lower cut thereby
defining a torus with two marked points corresponding to 
the infinities ${\rm Re}\,z\to\pm\infty$. The function
$G(z)$ is then uniquely specified by gluing
condition Eq.~\eqref{glue} and asymptotic behaviour at
large $\pm{\rm Re}\,z$
\EQ{\lim_{{\rm Re}\,z\rightarrow \pm\infty}G(z)\rightarrow\frac\mu
{4\sin(\tfrac{\beta m}2)}\,e^{\pm z}+{\cal O}(e^{\mp z})\ , 
\label{asymp}
}
which is a consequence of Eq.~\eqref{defg}.

The auxiliary torus 
$E_\ttau$ is specified by a complex structure $\ttau$ which can be
thought of as a function of the parameter $a$ specifying the length of the
cut. We can uniformize the torus by establishing a map to the complex 
$u$-plane ($u$ being an auxiliary variable) quotiented 
by a lattice with periods $2\pi i$ and $2\pi i\ttau$. 
The torus is shown in Figure 2, where for illustrative purposes
$\ttau$ has been taken to be purely imaginary.
As shown in
Figures 1 and 2, the contour $A$ enclosing the cut
$[-a+\tfrac{i\beta m}2,a+\tfrac{i\beta m}2]$
anti-clockwise maps to one of the cycles of the torus while
the contour $B$ joining the two cuts maps to the conjugate
cycle.  
\begin{figure}
\begin{center}\mbox{\epsfysize=5.5cm\epsfbox{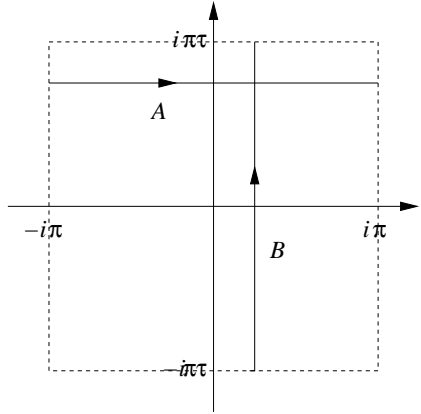}}\end{center}
\caption{\small Uniformizing map from the $z$-plane to the torus $E_\ttau$.}
\end{figure}

The map $z(u)$ from the $u$-plane to the $z$-plane is specified by the
requirements that going around the contour $A$ returns
$z$ to its original value, while traversing the contour
$B$ causes $z$ to jump by an
amount $i\beta m$, which is the distance between the two cuts in the
$z$-plane. Both these operations leave $G$ unchanged implying that it is
an elliptic function on the $u$-plane. Thus 
\AL{&A:\qquad z(u+2\pi i)=z(u)\ ;\qquad
\qquad\ G(z(u+2\pi i))=G(z(u))\ ,\\
&B:\qquad z(u+2\pi i\ttau)=z(u)+i\beta m\ ;\qquad
G(z(u+2\pi i\ttau))=G(z(u))\ .
\label{mono}
}
This determines the following unique map $z(u)$ from
the $u$-plane to the $z$-plane:
\EQ{
\exp{z(u)}=
{\theta_1(\tfrac{
u}{2 i}-\tfrac{\beta m}4\vert\ttau)\over\theta_1(\tfrac{
u}{2 i}+\tfrac{\beta m}4\vert\ttau)}\ .
\label{defz}
}
The only singularities of $G(z(u))$ are simple poles at
$u=\pm\tfrac{i\beta m}2$ corresponding to large $\pm{\rm Re}\,z$ in
order to incorporate the asymptotic behaviour \eqref{asymp}. The fact
that $G(z(u))$ is elliptic, along with the singularity structure, 
specifies it uniquely: 
\EQ{G(z(u))=\frac{i}{2\sin\tfrac{\beta
    m}2}\cdot\frac{\theta_1(\tfrac{\beta
    m}2|\ttau)}{\theta_1'(0|\ttau)}
\big(\zeta(u-\tfrac{i\beta}{2})-\zeta(u+\tfrac{i\beta}{2})
+2\zeta(i
\beta)-\tfrac\beta\pi\zeta(\pi i)\big)\ .
\label{degg}
}
In \eqref{defz} and \eqref{degg}, the quasi-elliptic function
$\zeta(u)$ is defined on the torus $E_\tau$ and $\theta_1(x|\tau)$ is
a Jacobi theta function 
(see \cite{WW} for definitions and Appendix A for some useful
identities). As expected from \eqref{mono},
$G(z(u))$ is an elliptic function of $u$ with two simple poles in
the $u$-plane; $z(u)$ on the other hand is only quasi-elliptic. 
Having determined $G(z)$ in the elliptic parameterization we
can now implement the Dijkgraaf-Vafa 
proposal in order to compute the superpotential in the
confining vacuum.

According the Dijkgraaf-Vafa proposal, the gluino condensate of the gauge
theory gets identified with the 't~Hooft coupling $S$ of the matrix
model. From Eqs.~\eqref{defres} and \eqref{defg}, the integral of  
$G(z)$ around the contour $A$ is equal to $-2\pi g_sN=-2\pi S$. Under
the map to the torus this becomes an integral around the $A$-cycle:
\EQ{
2\pi i S = -i\int _A G(z(u)){dz(u)\over du}du\ .
\label{wowa}
}
The second ingredient required to determine the QFT
superpotential is the variation of the genus zero free energy ${\cal
  F}_0$ of the
matrix model in transporting a test eigenvalue from infinity to the
endpoint of the cut. This is obtained by integrating the force on a
test eigenvalue, which
can be expressed in terms of the function $G(z)$ as 
\EQ{
-i\big(G(z+\tfrac{i\beta m}2)-G(z-\tfrac{i\beta m}2)\big)\ ,
} 
from infinity to the original cut $[-a,a]$. This can be written as an
integral over $G(z)$ alone along a contour starting at the 
lower cut,
going off to infinity and then back to the upper cut. This
can be deformed into the contour $B$ 
running from the lower cut to the upper
cut as in Figure 1. Under the map to the torus this becomes an
integral over the $B$-cycle:
\EQ{
{\partial {\cal F}_0\over \partial S}= -i\int _B 
G(z(u)){dz(u)\over du}du\ .
\label{dfds}
} 
The effective
superpotential in the confining vacuum is obtained by extremizing the following
expression with respect to $S$:
\EQ{
W_{\rm eff}=N{\partial {\cal F}_0\over \partial S}-2\pi i
\tau S\ ,
}
where $\tau$ is the bare coupling of the theory not to be confused
with the complex structure of $E_\ttau$; we shall shortly relate the
two. 

Both integrals \eqref{wowa} and \eqref{dfds} are 
evaluated using standard elliptic function identities:
\EQ{
2\pi i S= {d h(\ttau)\over d\ttau}\ ,\qquad \PD{{\cal F}_0}S=\ttau {d
h(\ttau)\over d\ttau}-h(\ttau)\ ,
}
where
\EQ{
h(\ttau)=\frac\mu{\sin\tfrac{\beta m}2}\cdot\frac{\theta_1(\tfrac{\beta
  m}2|\ttau)}{\theta_1'(0|\ttau)}\ .
\label{defh}
}
It follows that
\EQ{
\PD{W_{\rm eff}}S=0\quad  \Longrightarrow
\quad \ttau={\tau\over N}
}
so that
\EQ{
W_{\rm eff}=-Nh(\tfrac\tau N)=-\frac {N\mu}{\sin\tfrac{\beta
    m}2}\cdot\frac{\theta_1(\tfrac{\beta m}2|\tfrac\tau 
N)}{\theta_1'(0|\tfrac\tau N)}\ .
\label{res}
}

As a test of our expression for the superpotential 
we can consider the limit $\beta\to0$ and relate it to the known
result \cite{DV1,mm1,nick}. In this limit, from \eqref{res} we find
\EQ{
W_{\rm eff}\to -N\mu-\frac{N\mu m^2}{24}E_2(\tfrac\tau N)\beta^2+\cdots\ ,
}
where $E_2(\tau)$ is the second Eisenstein series. This result
agrees exactly with the expected result \cite{DV1,mm1,nick} 
since in this limit 
the potential behaves as $V(\Phi)\to\mu(1+\tfrac12\beta^2\Phi^2+\cdots)$.

It is possible to use the quantum mechanical matrix model to calculate
the superpotential in all the massive vacua of the theory. The idea is
to consider special multi-cut saddle-point solutions as described in
\cite{mm2}. We do not include the details here, but they can
straightforwardly be deduced from the aforementioned reference. 
The final result for the
$p^{\rm th}$ massive vacuum, where the confining corresponds to $p=1$
and the Higgs to $p=N$, is
\EQ{
W_{\rm eff}=-\frac {N\mu}{\sin\tfrac{\beta 
    m}2}\cdot\frac{\theta_1(\tfrac{p\beta m}2|\tfrac{p^2\tau} 
N)}{\theta_1'(0|\tfrac{p^2\tau}N)}\ .
\label{resv}
}
Recall that the result for the superpotential 
of the matrix model calculation in the
four-dimensional $\N=1^*$ case does not have exact modular
symmetry. For instance, the $S$ transformation
$\tau\to-1/\tau$ only relates the superpotential in the
Higgs and confining vacua up to an additive anomaly. In the
five-dimensional case described above we see that there is also a
modular anomaly in the matrix model result but now of a multiplicative
form. Interestingly under a general modular transformation 
\EQ{
\tau\to\frac{a\tau+b}{c\tau+d}
}
$i\beta m$ transformations like a point on the torus with complex
structure $\tau$:
\EQ{
i\beta m\to\frac{i\beta m}{c\tau+d}\ .
}

\section{The Elliptic Superpotential}

Another method that has been used to calculate the exact values of the
superpotential in all the massive 
vacua of the four-dimensional $\N=1^*$ theory is to
compactify it on a circle of finite radius \cite{nick,us}. The effective
superpotential is then a function of the dual photons and Wilson lines
of the abelian subgroup $\U(1)^{N-1}\subset\SU(N)$. These comprise
$N-1$ complex scalar fields $X_a$, $a=1,\ldots,N$ (with
$\sum_{a=1}^NX_a=0$) which naturally live on a torus of complex
structure $\tau$ because of the periodicity of each dual photon and
Wilson line. The superpotential 
describing the $\N=1^*$ deformation is therefore constrained to be an elliptic
function of the complex scalars $X_{a}$. It turns out that this
superpotential is identified as the basic Hamiltonian of the elliptic
Calogero-Moser integrable system where the $X_a$ are position
coordinates. The resulting superpotential, as
originally found in \cite{nick} can be 
\EQ{
W_{\rm elliptic}=\mu\Big(\sum_a\tfrac12P_a^2-
m^2\sum_{a\neq b}\wp(X_a-X_b)\Big)\ ,
\label{ees}
}
where $\wp(z)$ is the Weierstrass function defined on a torus with
periods $2\pi i$ and $2\pi i\tau$. Notice that the momenta can
trivially be integrated out. What is
particularly useful about this superpotential is that it is
independent of the compactification radius, and, therefore, yields
results that are valid in the four-dimensional limit. In addition
unlike the matrix model approach it encodes
all the vacua of the $\N=1^*$ theory, both massive and massless, in a
single superpotential.

The question is how this elliptic superpotential is generalized in the
five-dimensional theory compactified on a circle. In order to motivate
the answer we need to recall in more detail why the elliptic
Calogero-Moser integrable system appears in the basic $\N=1^*$ case. 
The reason is that, if
we think of the $\N=1^*$ theory in terms of a deformation by the mass
term $\mu$ of the
$\N=2^*$ theory, then the Coulomb branch of 
this latter theory is described by the moduli space of a
Seiberg-Witten curve. This curve is precisely the spectral curve of
the elliptic Calogero-Moser system \cite{donwitt}. Now for the
five-dimensional theory, Nekrasov has made a conjecture for the
Seiberg-Witten curve: it is precisely the spectral curve of a one
parameter deformation of the elliptic Calogero-Moser system known as
the elliptic Ruijsenaars-Schneider system. This is sometimes known as
the relativistic elliptic Calogero-Moser system.

Following the approach of \cite{nick,us} it is natural to conjecture
that the elliptic superpotential of the five-dimensional theory, for
the simplest deformation $\mu\cosh\beta\Phi$, is the basic Hamiltonian
of the Ruijsenaars-Schneider system. Based on this, and with reference
to \cite{Nek}, we expect
\EQ{
W_{\rm elliptic}=C\mu\sum_{a}\cosh(\beta P_a)\prod_{b\neq
  a}\sqrt{\wp(i\beta m)-\wp(X_a-X_b)}\ ,
\label{eesd}
}
for some constant $C$. It is easy to see that the critical points which
describe the confining vacuum of the $\N=1^*$ case
\EQ{
P_a=0\ ,\qquad X_a=\frac{2\pi ia\tau}N\ ,\qquad a=1,\ldots,N\ ,
}
are still critical points of the superpotential \eqref{eesd}. In fact
this is true for all the massive vacua described in \cite{nick}.

We can now compare the result of the quantum mechanical matrix model
\eqref{res} with the result from the elliptic superpotential:
\EQ{
W_{\rm elliptic}=
NC\mu\prod_{a=1}^{N-1}\sqrt{\wp(i\beta m)-\wp(\tfrac{2\pi ia\tau}N)}\ .
}
Amazingly the results agree by virtue of the elliptic function
identity \eqref{magic} proved in the Appendix
\EQ{
\frac{\theta_1(\tfrac\beta2|\tfrac\tau N)}{\theta_1'(0|\tfrac\tau N)}
=(2i)^{N-1}
\Big(\frac{\theta_1(\tfrac\beta2|\tau)}{\theta_1'(0|\tau)}\Big)^N
\prod_{a=1}^{N-1}\sqrt{\wp(i\beta m|\tau)-\wp(\tfrac{2\pi ia\tau}N|\tau)}\ ,
}
where we have emphasized the complex structure
associated to the elliptic functions on the right-hand side.

\startappendix

\Appendix{Some Properties of Elliptic Functions}

We use the (quasi-)elliptic functions $\wp(u)$, $\zeta(u)$,
$\theta_1(\tfrac u{2i}|\tau)$ associated to a torus of periods $2\omega_1=2\pi
i$ and $2\omega_2=2\pi i\tau$ as defined in \cite{WW}. An important
equation relating them is
\EQ{
\zeta(u)-\frac{\zeta(\omega_1)}{\omega_1}u=\frac{1}{2i}
\frac{\theta'_1(\tfrac u{2i}|\tau)}
{\theta_1(\tfrac u{2i}|\tau)}\ ,
\label{relt}
} 
We also use the heat equation
\EQ{
\frac{\partial^2\theta_i(x|\tau)}{\partial x^2}+
\frac4{\pi i}\frac{\partial\theta_i(x|\tau)}{\partial\tau}=0
\label{heat}
}
and the relations 
\EQ{
\zeta(\omega_1)\omega_1=\frac{\pi^2}{12}\cdot E_2(\tau)=-
\frac{\pi^2}{12}\cdot\frac{\theta_1^{\prime\prime\prime}(0|\tau)}
{\theta_1'(0|\tau)}\ .
\label{url}
}

In the remainder of the Appendix, we establish a crucial identity. We
start with the following identity established in 
\cite{cft} (using \eqref{relt}, \eqref{url} and other standard 
properties of elliptic functions) 
\EQ{
{\theta_1^\prime(\tfrac u{2i}|\tfrac\tau N)\over
    \theta_1(\tfrac{u}{2i}
\vert\tfrac\tau N)}-{\theta_1^\prime(\tfrac{u}{2i}|\tau)\over \theta_1(
\tfrac{u}{2i}
\vert\tau)}+\frac{iu(N-1)}6E_2(\tau)
=i\sum_{a=1}^{N-1}\big(\zeta(\tfrac{2\pi ai\tau}N+u
)-\zeta(\tfrac{2\pi ai\tau}N-u)\big)\ .
}
We now integrating this expression, using the relation \eqref{relt} on
the right-hand side, to arrive at
\EQ{
\frac{\theta_1(\tfrac{u}{2i}|\tfrac\tau N)}{\theta_1(\tfrac{u
    }{2i}|\tau)}=\prod_{a=1}^{N-1}\sqrt{\theta_1(\tfrac {\pi a\tau}N-
\tfrac{u}{2i}|\tau)\theta_1(\tfrac{\pi a\tau}N+\tfrac{u
  }{2i}|\tau)}\ .
}
Further, on the right-hand side, we can employ the relation
\EQ{
\frac{\theta_1(\tfrac {u-v}{2i}|\tau)\theta_1(\tfrac
  {u+v}{2i}|\tau)}{\theta_1(\tfrac
  u{2i}|\tau)^2\theta_1(\tfrac v{2i}|\tau)^2}=\wp(v)-\wp(u)
}
to derive the identity
\EQ{
\theta_1(\tfrac u{2i}|\tfrac\tau N)=\theta_1(\tfrac
u{2i}|\tau)^N\prod_{a=1}^{N-1}\theta_1(\tfrac{\pi a \tau}N|\tau)\sqrt{
\wp(u)-\wp(\tfrac{2\pi ia\tau}N)}\ .
}
From the ratio of this, and its derivative evaluated at $u=0$, 
we have our crucial identity
\EQ{
\frac{\theta_1(\tfrac u{2i}|\tfrac\tau N)}{\theta'_1(0|\tfrac\tau N)}
=(2i)^{N-1}\Big(\frac{\theta_1(\tfrac u{2i}|\tau)}
{\theta'_1(0|\tau)}\Big)^N\prod_{a=1}^{N-1}\sqrt{
\wp(u)-\wp(\tfrac{2\pi ia\tau}N)}\ .
\label{magic}
}

\end{document}